\documentclass[sigconf]{acmart}
\usepackage{booktabs}
\usepackage{multirow}
\usepackage{longtable}

\newcommand{\fix}[1]{}

\newcommand{\rev}[1]{#1}

\definecolor{chestnut}{rgb}{0.8, 0.36, 0.36}
\newcommand\rk[1]{}

\AtBeginDocument{%
  \providecommand\BibTeX{{%
    \normalfont B\kern-0.5em{\scshape i\kern-0.25em b}\kern-0.8em\TeX}}}

\acmConference[MSR '23]{20th International Conference on Mining Software Repositories}{May 15--16, 2023}{Melbourne, Australia}
\begin{document}

\title{Mining the Characteristics of Jupyter Notebooks\\ in Data Science Projects }

\author{
% Anonymous Authors
Morakot Choetkiertikul, 
Apirak Hoonlor, 
Chaiyong Ragkhitwetsagul, 
Siripen Pongpaichet, 
Thanwadee Sunetnanta, 
Tasha Settewong, 
Vacharavich Jiravatvanich,
Urisayar Kaewpichai
}
\email{{morakot.cho, apirak.hoo, chaiyong.rag}@mahidol.ac.th}
\email{{siripen.pon, thanwadee.sun}@mahidol.ac.th}
\email{{tasha.set, vacharavich.jir, urisayar.kae}@student.mahidol.ac.th}
\affiliation{%
 \institution{Faculty of ICT, Mahidol University}
 \city{Nakhon Pathom}
 \country{Thailand}
}

\author{
Raula Gaikovina Kula
}
\email{raula-k@is.naist.jp}
\affiliation{%
 \institution{Nara Institute of Science and Technology (NAIST)}
 \city{Nara}
 \country{Japan}
}

\renewcommand{\shortauthors}{Choetkiertikul et al.}

\begin{abstract}

Nowadays, numerous industries have exceptional demand for skills in data science, such as data analysis, data mining, and machine learning. The computational notebook (e.g., Jupyter Notebook) is a well-known data science tool adopted in practice. Kaggle and GitHub are two platforms where data science communities are used for knowledge-sharing, skill-practicing, and collaboration. While tutorials and guidelines for novice data science are available on both platforms, there is a low number of Jupyter Notebooks that received high numbers of votes from the community. The high-voted notebook is considered well-documented, easy to understand, and applies the best data science and software engineering practices. \rk{Is better if we form a hypothesis and then work to prove this. This will provide more direction} In this research, we aim to understand the characteristics of high-voted Jupyter Notebooks on Kaggle and the popular Jupyter Notebooks for data science projects on GitHub. We plan to mine and analyse the Jupyter Notebooks on both platforms. We will perform exploratory analytics, data visualization, and feature importances to understand the overall structure of these notebooks and to identify common patterns and best-practice features separating the low-voted and high-voted notebooks. Upon the completion of this research, the discovered insights can be applied as training guidelines for aspiring data scientists and machine learning practitioners looking to improve their performance from novice ranking Jupyter Notebook on Kaggle to a deployable project on GitHub.

\end{abstract}






\maketitle

\section{Introduction}

% At present, data analysis, data mining, and machine learning is in high demand because of the integration of feasibility of a wide range of applications in various industries \cite{mitchell1997machine}. The relevant organization has been established to respond to the increasing demand for specialized knowledge in data analysis which provided providing computational notebooks that assist data scientists in data exploration. 

According to \cite{Medeiros}, a company with an investment in big data or data science has shown an increase in productivity from $3\%$ to $7\%$. \fix{R:R2.12} \rev{Over the past five years, we have observed an increase in various data-driven application deployments to drive business in small-, to medium-, sized organizations.} The integration of software engineering, data analysis, data mining, and machine learning techniques has brought about a high demand in various industries \cite{10.1145/2500499, provost2013data, van2016process}. Such integration increases collaboration between data scientists and software engineering in the development teams~\cite{Kim2016}. Due to the difference in nature of the data science project in comparison with those of a software project, the development teams must adjust their practice to the requirements of data science projects. In addition, data analysis, data mining, and machine learning techniques have been applied to improve software development and software maintenance processes \cite{ZHANG20167}. To train data science skills, one must practice in various data science projects \cite{SEDS,CMU}. GitHub and \rev{Kaggle\footnote{\url{https://www.kaggle.com/}}} are two prominent data-science communities on their platforms, which offer data science training resources as well as data science projects for practicing data science skills. 

\fix{R:R3.2}
\rev{On GitHub, such resources are offered via multiple projects ranging from training computational notebooks to a fully deployed data analytic project. On Kaggle, a cloud-based collaborative platform, practitioners conduct and contribute computational notebooks for machine learning competitions on various topics. The  common resource for training is a computational notebook. A computational notebook, such as Jupyter Notebook, is a web-based interactive computing environment with executable codes. It allows a user to implement codes, see the results, and provide discussions on the computational notebook results in easier-to-understand data science work. This popularizes the computational notebook as a tool for not only reproducing, but also for tutoring and training purposes.} In addition to hosting the collections of computational notebooks, Kaggle also hosts both rewarded and non-rewarded competitions, providing opportunities for individuals or teams to compete or contribute notebooks. Hence, one can expect the computational notebook shared via a project on GitHub to vary from those on Kaggle's competitions.\fix{R:R2.12} \rev{Since both platforms offer resources ranging from tutorial notebooks to top-ranking competition notebooks, some of the notebooks are challenging for a newcomer to understand and learn how to master the art of data science.} 

Kaggle recognizes this problem and helps the beginner by providing courses and guides (a curated list of high-quality resources). Some of the projects on GitHub also provide learning and training resources for beginners. However, only a small portion of community members are considered grandmasters on Kaggle. In addition, the shared data science-related computational notebooks are not always implemented with software engineering practices such as maintainability and readability. To this end, we want to aid and identify key features to help practitioners on Kaggle to increase their data science skills, as shown through the quality of their computational notebooks to the level of public data science projects on GitHub. We begin our work on this project by investigating data science projects on both platforms.

For Kaggle, it guides the development of data science practitioners from beginner to expert by having a contribution level of the notebook to identify the high-quality contribution. Kaggle measures the contribution level of the notebook based on the number of votes from experienced practitioners. In its contribution ranking, Kaggle rewards contributors with medals and higher tiers, which reflect consistent and high-quality contributions. Becoming a grandmaster on Kaggle can be a challenging task, especially for newcomers who may face various obstacles in their practice. One such obstacle is the time required to gauge the impact of a notebook, as it takes time to receive upvotes from other users and determine its medal in a competition. Additionally, the process of becoming a notebook grandmaster, who is awarded the highest tier through consistent contributions of high-quality notebooks, requires not only expertise in skills but also a deep understanding and perspective in the field of data science. To the best of our knowledge, there is currently no research on identifying the characteristics of grandmaster-tier notebooks in Kaggle. To help newcomers overcome these challenges, we aim to study the factors that influence the progression of notebooks, with a focus on the characteristics of notebooks created by users who have achieved the grandmaster tier. By identifying these characteristics, we can classify notebooks based on their level of expertise (e.g., novice or grandmaster), which will aid newcomers in determining the quality of notebooks. 

For GitHub, we will identify and study its public data science projects. We will analyse those projects' code qualities and the characteristics similar to those found in the grandmaster notebooks on Kaggle. These will give us an insight into the difference between the data science projects on Kaggle and those on GitHub. To this end, we hope to provide suggestions to improve notebook quality and guidelines to further develop a project from a computational notebook to a deployable project. In summary, given the challenges faced by newcomers in the field of data science and machine learning, this research aims to:

\fix{R:R2.3}
\rev{\begin{itemize}
    \item Identify and extract key characteristics of the computational notebooks from Kaggle and GitHub.
    \item Investigate the factors that influence the quality and success of these notebooks.
\end{itemize}}

The rest of this registered report is organized as follows. In the next section, we provide the background and related work. We define and explain our research questions in Section \ref{RQ}. Our plan of execution is detailed in Section \ref{plan}. We discuss the implication of our work in Section \ref{section:impact} and conclude our current work in Section \ref{conclusion}.

\section{Background and related work}
In this section, we explain the background of our study, including the description of a computational notebook, Kaggle, GitHub, and related work. 

% web-based interactive computing application for contributing and cooperating the notebook in various programming languages, which combines with two components (A web application and a notebook document)
% 1. web application: {explaination + feature}
% 2. notebook document: {explaination + feature}

\subsection{Computational notebook}
A computational notebook, such as Jupyter Notebook and R Markdown, is a coding platform that enables the combination of written text and executable code, with the results of the code being incorporated into the document \cite{Hara2015}. The notebook document is visible content in the web application associated with the inputs and outputs of the computations, explanatory text, mathematics, images, and rich media representations of objects. There are typically three types of contents, referred to as ``cells'', which include code cells, markdown cells, and raw cells. Code cells contain executable code (e.g., Python), markdown cells contain texts and formatting, and raw cells contain unformatted texts. These cells can be run independently or together to perform computations, analyse data and write results in a structured and readable manner. Such a coding platform eases the process of reproducibility of analytical tasks and sharing analytical codes. For instance, Jupyter Notebook provides a web-based interactive computing environment that enables users to create a notebook for conducting computations, making it a valuable tool for data analysis and manipulation. According to \cite{Jeffrey2018}, Jupyter is considered the tool choice for data science tasks, with over $2.5$ million public Jupyter Notebooks found on GitHub alone. 

%\footnote{\url{https://jupyter.org/}} should we add R markdown too? either we have both or not include them... I think we have reference?

% The components of Jupyter Notebook consist of a web application and a notebook document. The web application is an interactive tool associated the explanatory text, mathematics, computations, and rich media output. 

% that allows users to contribute to the notebook or cooperate with other users. This web application enables the user to conduct the notebook with over a hundred programming languages that offer a simple, streamlined, document-centric experience \cite{jpynb1}. 

% The code cell is sent to the kernel associated with the notebook when it is executed. The result from executing is then displayed as the cell’s output which is not limited only to text. The output can also be displayed in many other possible forms, such as matplotlib figures and HTML tables, known as IPython’s rich display capability. A markdown cell allows the user to use descriptive text with code as a literate way to document the computational process. In Ipython, the markdown language is used to mark up text. The corresponding cells are called Markdown cells. The Markdown language provides a simple approach to perform this text markup. Within the markdown cell, users can specify the part they want to emphasize using an italic, bold, and form list. Structuring the document can also be done by using a markdown heading. A raw cell is a space that allows the user to write output directly since it is not evaluated by the notebook.

\subsection{Kaggle}
% Kaggle\footnote{\url{https://www.kaggle.com/}} is a cloud-based collaborative platform for practitioners to learn data science, a practitioner with experience with data science, and a practitioner with expertise in the field of data science, specifically machine learning. The platform provides multiple tools and resources to data scientists. 

Kaggle is a cloud-based computational notebook platform that serves as a collaborative space for individuals interested in data science. The platform offers a wide range of tools and resources for data scientists, including access to datasets, competitions, and a community of users to share knowledge and collaborate with. Especially, Kaggle also offers a platform for individuals to build their profile in the field of data science. This can be achieved through participation in various 
activities such as analyzing datasets, participating in competitions, and contributing to the Kaggle community. Through these activities, users can gain visibility and demonstrate their skills and knowledge in data science practices. For example, Kaggle hosts a competition that challenges participants to develop a model for predicting loan defaults using a dataset provided by American Express, with a prize of 100,000 USD awarded to the winner.\footnote{\url{https://www.kaggle.com/competitions/amex-default-prediction}} 

As mentioned in the Introduction, apart from rewards from competitions, Kaggle has a ranking system to reflect the contributions of its users. This ranking system consists of different levels, which are achieved through participation in competitions, contributing to the community, and other activities on the platform. The progression through the ranks serves as a way for users to showcase their skills and experience. In this ranking system, a notebook's rank is determined by the number of upvotes received from other users. Thus, users who consistently produce high-quality work have the potential to achieve higher ranks and make progress toward their ranking within the platform. This incentivises users to produce high-quality work and encourages them to share their knowledge and collaborate with others in the community. There are five ranks in Kaggle: Novice, Contributor, Expert, Master, and Grandmaster. A Novice is considered a user who is new to Kaggle, and who just started to learn and explore the platform. In contrast, Grandmaster is the highest tier in Kaggle's ranking system and the most respected rank in the community. This rank is awarded to users who have consistently produced exceptional work. To reach the Grandmaster tier, a user typically needs to have a certain number of medals, which are awarded for various contributions such as winning competitive notebooks, writing high-quality notebooks, or participating in challenging notebooks. \fix{R:R1.10 fixed}For example, a notebook with five upvotes is awarded a \rev{Bronze medal}, a notebook with twenty upvotes is awarded a Silver medal and a notebook with fifty upvotes is awarded a Gold medal. The Grandmaster tier can be achieved by obtaining 15 Gold medals. These incentives thus serve as a way to recognize and acknowledge the high-quality work of the users and their contributions to the community.

\subsection{GitHub}

GitHub provides a version control system for software development and a hosting service for software projects. For data science projects, GitHub can be used to store and share code, data, and documentation in the form of computational notebook files such as Jupyter Notebook files. The ability to store and share computational notebooks on GitHub enables team members to collaborate and track the progress of data science projects. GitHub also contains multiple types of training materials in data science and related areas ranging from a curated list of free books \cite{BrendanMartin} to tutorial projects \cite{donnemartin}. 
%Apirak will add a bit more example of tutorial page and data science project on GitHub

\subsection{Related work}

% \subsection{Studying of Kaggle Platform}
%study on kaggle data, explain 

\rev{Wang et al. \cite{Wang2021} provide a contribution to the understanding of data science documentation practices on Kaggle based on the textual description in markdown cells and code cells}. By considering highly-voted notebooks as a proxy for well-documented notebooks, the authors are able to gain insights into what makes a notebook well-received by the Kaggle community. According to the preliminary findings of the study, there appears to be a difference between the top-voted notebooks by the Kaggle community and the top-ranked ones on the competition leaderboard, which is determined by the performance metrics solution and the competition's objective, specifically the accuracy of the model's predictions. This suggests that the Kaggle community places a significant value on factors beyond performance, such as clear documentation, reproducibility, and ease of use. Based on their findings, the authors of the study formulated the hypothesis that the high-voted notebooks receive a high number of upvotes due to their high levels of readability and comprehensive documentation, which may have contributed to their popularity among the Kaggle community. The studying of 80 high-voted notebooks selected from the top 1\% of all notebooks submitted in the two most popular Kaggle competitions shows that the textual description in those notebooks provided a comprehensive description of the code cells. However, our work focuses on a broader range of notebook characteristics (e.g., code quality) and incorporates data from GitHub, which provides a more comprehensive understanding of the factors that contribute to the success of a notebook in data science. 

 % They conducted a qualitative analysis on 80 highly-voted notebooks collected from the top 1\% of the notebooks submitted in the two most popular competitions. As a result of their qualitative analysis, the notebooks that received high upvotes on Kaggle were well-documented. The content in markdown cells covered a broad range of topics and purposes to describe the adjacent code cell.

% From their preliminary investigation that the top-voted notebooks by community members were not likely to be in the top-ranked ones on the leaderboard, which is considered based on the performance metrics solution (i.e., the accuracy of the result predicted by the model), they came up with the hypothesis that these highly-voted notebooks obtain the upvotes as a consequence of high readability and better document.

% \subsection{Bug bounty program}

Furthermore, our work also relates to previous empirical studies that have investigated platforms that provide rewards or incentives to participants, such as monetary compensation or a reputation score, since the user ranking advancement system in Kaggle can be compared to bug bounty programs, as both offer incentives for contributors. Walshe et al. \cite{9034828} conducted a study on the bug bounty or vulnerability reward program (VRP), which involves offering rewards to white hat hackers to locate and report vulnerabilities in software. The VRP approach has become increasingly popular as a way to identify security flaws in software systems. The study reports that the amount of monetary reward offered did not influence the number of vulnerability reports submitted by the hackers. Instead, they found that the hackers were motivated by their reputation posted on the websites where they participated. In addition, the study by Kanda et al. \cite{Kanda2017} found that projects with bounties on the Bountysource platform are more likely to be solved than those without bounties. Further research by Zhou et al. \cite{Zhou2020, Zhou2021} showed that the bounty value is not the most significant factor that attracts contributors to work on the issues, as some contributors may be motivated by their own interests or desires rather than solely by rewards or monetisation.

\section{Research questions}
\label{RQ}
As discussed in the introduction, our goals are to aid and identify key features to help practitioners on Kaggle increase their data science skills. Our first step toward this goal is investigating data science projects on both platforms. \fix{R:R1.3} \rev{Specifically, the quality of data science projects on Kaggle and GitHub can vary widely due to the diverse range of projects and contributors with varying levels of skill and experience, both platforms offer valuable resources for data science practitioners to learn from and improve their skills. By comparing the characteristics of computational notebooks between these two collaborative platforms, practitioners can gain insight into the notebook's characteristics reflecting the practices for organizing and presenting data science work.}
% Specifically, we want to identify the features that the data science practitioners should improve on based on their current skills in order to increase the quality of their notebooks to that of the public data science projects on GitHub.
\rk{we can posit that the quality of the notebook increases over time}
To this end, we define three following exploratory research questions.

\begin{itemize}
    \item \textbf{RQ1}:~\textit{What are the important features determining the ranking of notebook contributors?}
    \item \textbf{RQ2}:~\textit{Do the changing of notebook characteristics correspond to the notebook contributor ranking progression?}
    \item \textbf{RQ3}:~\textit{Do the characteristics of grand master's notebooks correlate with the highly popular notebooks in GitHub projects?} 
\end{itemize}

To address the first two research questions, we will identify and extract features from Jupyter Notebook on the Kaggle platform. Then, we will track the changes in these features as the contributors improved from novices to grand masters. For the third research question, we plan to investigate whether findings can be applied to open-source data science projects using Jupyter Notebooks outside of Kaggle's context. We will sample data science projects from GitHub containing Jupyter Notebooks. Then, we will repeat the process in RQ1 and study the correlation by using a correlation test that matches the data. We provide the full details of our plans, including the list of targeting features and GitHub project selection criteria, in the next section. 

\section{Execution plan}
\label{plan}

\begin{figure*}[htbp]
    \centerline{\includegraphics[width=\textwidth]{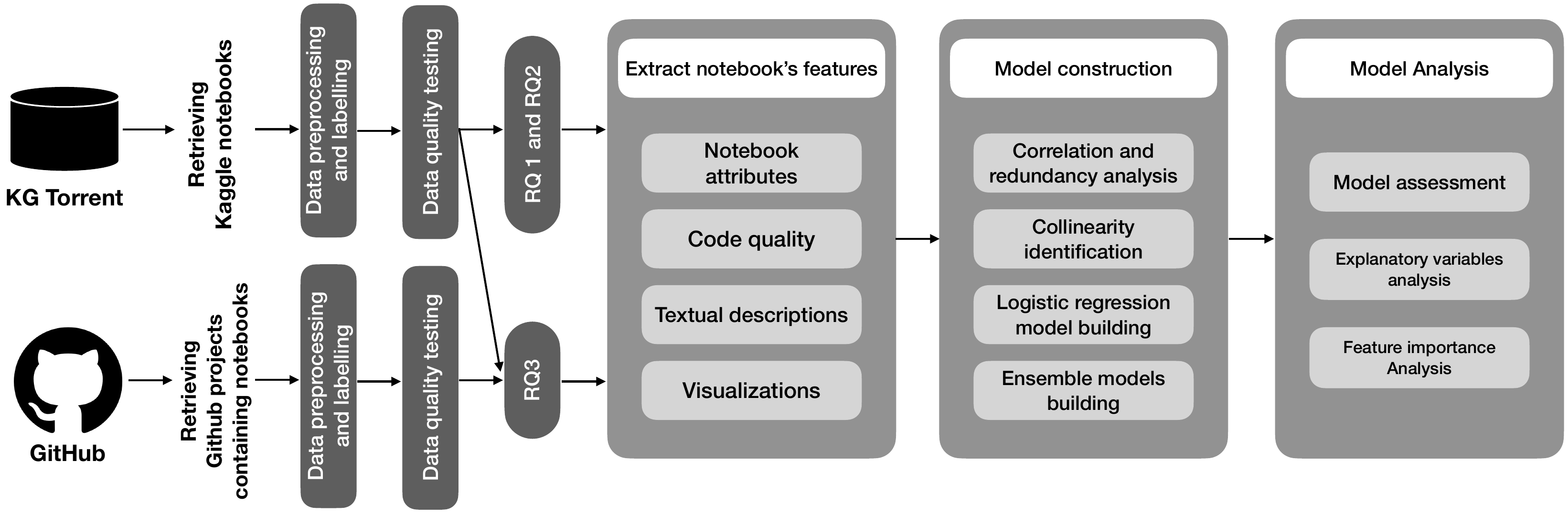}}
    \caption{Overview of our exploratory study \fix{R:R2.6}}
    \label{fig:methodology}
\end{figure*}

\fix{R:R2.6}
In this section, we outline our approach to executing our research. We have chosen an exploratory case study design that adheres to ACM guidelines for the experiment \cite{acmguide}. Figure \ref{fig:methodology} presents an overview of our research study to address and answer our research questions. In this research execution plan, the first phase focuses on the collection, preprocessing, and labelling of the data. \fix{R:R2.6} \rev{We test our data quality two ways. For the preprocessed data, we create the SQL assertion tests. For the labelling, we performed manual validation.} In the subsequent phase, we will extract the features that characterize computational notebooks. This extraction process involves four distinct groups of features, including general notebook attributes, features related to textual descriptions, features related to visualizations, and features related to code quality. The next step is to construct models and conduct the analysis. The models will be designed to answer the research questions and test the hypothesis. The analysis will be performed using various statistical techniques and machine learning techniques, such as regression analysis and random forests. 
% -- Stopped here

\subsection{Data collection}

Our research plan involves the examination of computational notebooks data from two sources: Kaggle and GitHub. For Kaggle, we are using the KGtorrent dataset \cite{9463068}, while the data from GitHub will be collected by ourselves. KGTorrent \cite{9463068} is a comprehensive dataset that consists of computational notebooks (i.e., code kernels) with accompanying metadata collected from Kaggle. The metadata contained in KGTorrent includes information about the notebooks, user profile achievements, and details about the competitions hosted on Kaggle. In addition, to gather computational notebook data from GitHub, we will collect open-source projects containing computational notebook files (e.g., Jupiter Notebook files) that meet our criteria through GitHub's API. 

% \subsubsection{KGTorrent}
% dataset characteristics, meta data
% KGTorrent \cite{9463068}, our selected data source, is a dataset containing 248,761 Python Jupyter notebooks with rich metadata retrieved from the Meta Kaggle, which is the dataset that is available in CSV files through Kaggle kernels, and containing tables on public activity from Competitions, Datasets, Kernels, Discussions. Therefore, KGTorrent leverage Meta Kaggle and build a companion MySQL database recording those metadata indicating notebooks, users, and competitions, e.g., code kernels, user’s achievements for over five million Kaggle users, and over two thousand competitions.

\subsection{Data preprocessing and labelling}
% October 2020 is the final month of data on KGTorrent
\subsubsection{KGTorrent dataset} In order to use the data from KGTorrent \cite{9463068}, we have investigated the dataset and performed a filtering process as the following criteria: 1) the notebooks must be from closed competitions, 2) the notebook's metadata and \rev{its}\fix{R:R2.13} corresponding notebook files must \fix{R:R2.14} both exist in the dataset, and 3) the information of a contributor who created a notebook must be available in the dataset. Table \ref{table:kgtorrentdataset} shows the number of notebooks that passed our filter in each tier based on the data recorded in the Kaggle meta data (i.e., user profile). In total, we can retrieve 11,939 notebooks. %for our study. 

In this study, the notebook creator tiers are used as a proxy to indicate the quality of the computational notebooks, as the progression and achievement of the creator tier are representatives of community recognition and appreciation. These notebooks will be labelled as tier levels from 0 to 4 \fix{R:R2.4} (i.e., \rev{target variable}) based on the classification used in the Meta Kaggle database: 0 represents the rank of a novice, 1 represents a contributor, 2 represents an expert, 3 represents a master, and 4 represents a grandmaster. However, it is important to note that using the creator tier recorded in the metadata may not be entirely accurate, as the rank of the creator may have changed at the time the notebook was created. We will then perform notebook labelling by considering the progression of the user over time. In this method, we chronologically categorize the notebooks produced by each user and label them based on their tier criteria by processing the user's profile changelog recorded in the KGTorrent dataset. For example, a notebook is labelled as ``grandmaster'' tier if the creator had accumulated at least 15 gold medals at the time of creating the notebook. \rev{We then incorporate a correlation analysis between contributor rank and notebook up-votes and a manual validation on labelling in our study. This analysis can provide valuable insights into the relationship between these two variables and help to validate the appropriateness of using the extracted contributor rank.} 

% -- Stopped here

\rev{We acknowledge that the data is imbalanced. To address this concern, we aim to apply various techniques, including resampling methods and stratified sampling during model creation. Additionally, we will employ machine learning algorithms that allow for adjusting class weights, which can help account for the data imbalance. Furthermore, we will focus on evaluation metrics that are robust to class imbalance, ensuring a more reliable assessment of our model's performance.} 
\fix{R:R3.4}\fix{R:R1.4} \fix{R:R3.3}

\begin{table}[h]
\centering
\begin{tabular}{lrr}
\hline
\textbf{Total number of}                            & \textbf{Numbers}  & \textbf{Percentage} \\ \hline
Notebooks in KGTorrent                              & 248,761           &                     \\
Retrieved notebooks\hspace{0.5cm}    & 11,939            &                     \\
\hspace{0.5cm}Novice                                & 6,591             & 55.21\%             \\
\hspace{0.5cm}Contributor                           & 3,388             & 28.38\%             \\
\hspace{0.5cm}Expert                                & 1,674             & 14.02\%             \\
\hspace{0.5cm}Master                                & 215               & 1.80\%              \\
\hspace{0.5cm}Grandmaster                           & 71                & 0.59\%              \\ \hline
\end{tabular}
\caption{Notebooks retrieved from KGTorrent}
\label{table:kgtorrentdataset}
\end{table}

\begin{table*}[]
    \centering
    \caption{Features identified and extracted from Jupyter notebooks \rev{(* indicates that this feature was also used in \cite{Wang2021})} \fix{R:R1.8} \fix{R:R1.9} \fix{R:R2.8}}
    \resizebox{0.9\textwidth}{!}{%
    \begin{tabular}{@{}p{3cm}lp{7cm}p{5cm}@{}}
    \toprule
    \textbf{Group} &
      \textbf{Feature Name} &
      \textbf{Description} &
      \textbf{Rationale} \\ \midrule
    Notebook attributes &
      Dataset &
      Whether the dataset used in a notebook is the same one as the dataset provided in the competition &
      \multirow[t]{5}{5cm}{The alignment between   the usage dataset, notebook's tags, and competition tags may describe an   objective of a notebook. Good quality notebooks may have a strong alignment   among these variables} \\
     &
      Notebook tag &
      List of tags of a   notebook &
       \\
     &
      Competition tag &
      List of tags of a   competition asscociated with a notebook &
       \\
       & Cosine sim tags & The cosine similarity score between Notebook's tags and Competition's tags & \\ 
       \midrule
    Code quality &
      Number of code cells\rev{*} &
      The number of code cells in a notebook &
      \multirow[t]{15}{5cm}{For data science, the code quality affects the reproducibility and verifiability of the results in notebooks. For software engineering, code quality affects the understanding and maintainability of the software and encourages collaboration which is a significant concern in software engineering. Overall, higher code quality can lead to better sources for training. \rev{Note that some features are at a lower level (e.g., Cyclomatic complexity) while others are at a higher level (e.g., Maintainability) as these metrics reflect different aspects of software quality.}} \\
     &
      Number of code lines\rev{*} &
      The total number of   code lines in a notebook &
       \\
     &
      Number of comment   lines &
      The total number of   code comment lines in a notebook &
       \\
     &
      AVG. code lines per   cell &
      The mean of numbers   of code lines per code cell &
       \\
     &
      Number of functions &
      The number of   functions declared in a notebook &
       \\
     &
      Cyclomatic complexity &
      \rev{The score reflecting the cyclomatic complexity of the code calculated based on the number of paths through the code}&
       \\
     &
      Cognitive complexity &
      \rev{The score reflecting the understandability of code based on the code's control flow and considers factors such as nesting, branching, and the use of logical operators, which contribute to the difficulty of comprehending the code} &
       \\
     &
      Duplication blocks &
      The number of   duplicated code blocks of lines in a notebook &
       \\
     &
      Duplication lines &
      The number of   duplicated code lines in a notebook &
       \\
     &
      Code smell &
      The number of   identified code smell issues &
       \\
     &
      Technical debt &
      The amount of effort   required to fix all code smells &
       \\
     &
      Maintainability &
    \rev{The score reflecting the maintainability of code based on the code's Technical Debt Ratio and the number of lines of code} &
       \\
     &
      Reliability &
      \rev{The score reflecting   the reliability of code based on  the number of bugs or defects per unit of code (Bug density)}   &
       \\
     &
      Vulnerability &
      The number of   vulnerability issues &
       \\
     &
      Security &
      The score reflecting   the security rating of code &
       \\\\ \midrule
    Textual descriptions &
      Number of markdown   cells\rev{*} &
      The number of   markdown cells in a notebook &
      \multirow[t]{6}{5cm}{Readability metrics and other textual-related features can help better understand notebooks.
      The initial assessment from grandmaster notebooks, the quantity and quality of textual description are much higher than those of novice notebooks.} \\
     &
      Number of markdown   lines &
      The number of lines   in markdown cells in a notebook &
       \\
     &
      Flesch score &
      The readability   metric computed from sentence length and syllables/words &
       \\
     &
      Number of sections &
      The number of headers   in markdown cells in a notebook &
       \\
     &
      Avg. sentences per   cell &
      The mean of numbers   of sentences per markdown cell &
       \\
       & & & \\ 
       \midrule
    Visualizations &
      Number of   visualizations &
      The number of   visualizations used for visualizing data &
      \multirow[t]{4}{5cm}{According to Settewong et al.~\cite{Settewong2022}, there are different categories of visualizations used in notebooks. Visualizations are tools to communicate the understanding and insights of the data.} \\
     &
      Number of imported   images &
      The number of   imported images and figures in a notebook &
       \\
     &
      Visualization   libraries &
      List the imported   library used in visualizations &
       \\
     &
      Visualization   functions &
      List of functions   used for creating visualizations e.g. bar, line &
       \\ \\\bottomrule
    \end{tabular}%
    }
    \label{tab:features}
    \end{table*}
\subsubsection{GitHub data}
\label{sec:github} 
% -- Stopped here

For our study, we will collect open-source data science projects hosted on GitHub using their API. The projects must satisfy the following criteria: 1) the project must have at least ten stars to filter out trivial or toy projects, and \rev{2) the projects must contain Jupyter Notebooks as the majority of all files. For example, the ratio of Jupyter Notebook files must be greater than 60\%. This cutoff threshold will be determined based on the distribution of the Jupyter Notebook file ratios in the selected projects.} After obtaining the initial list of projects, we will use the number of stars and forks as proxies \fix{R:R2.4} (i.e., \rev{target variable}), as opposed to using contributor ranks in the Kaggle notebooks. The numbers of stars and the number of forks reflect popularity as perceived by the GitHub community\rev{~\cite{Borges2016}}, which is similar to the ranking mechanisms in Kaggle.  \fix{R:R2.7} \rev{For the number of stars, we followed the same criteria suggested in~\cite{Dabic2021}, where the authors suggested ten stars as a good compromise between the quality of data and the time required to mine the GitHub projects.}
\rev{We will control the quality of our GitHub project sampling as follows. First, to mitigate the threats to validity that the star is given to a GitHub project and may not fully relate  to the Jupyter Notebook files, we have set our selection criteria to only projects that contain Jupyter Notebook files as the majority compared to all other files. Second, we will apply the technique presented in Munia et al.'s study~\cite{Munaiah2017} to remove toy projects or tutorials and keep only engineered software projects.}

\subsection{Features}

\fix{R:R2.8} The summary of relevant features for all research questions is shown in Table \ref{tab:features}. We will extract four groups of features: 1) the notebook features indicate the basic information of a notebook e.g., notebook's tags, 2) the code quality metrics describe the features of the code cells contained in the notebook, e.g., code complexity, 3) The textual description-related features capture the characteristics of the markdown cells, e.g., readability score, and 4) the visualization related features reflect the use of data visualization techniques appeared in the notebook, e.g., the number of visualizations.

 % \subsubsection{Correlation attribute}
 % The correlation between the two attributes is reflected in this group. For instance, an average line of code per cell, an average char per cell, and the ratio of markdown cell to code cell.

\subsection{Analysis method}
For each research question, we explain our plan to analyse the data using statistical testing, feature importance, and correlation below.

\textbf{RQ1:~What are the important features determining the ranking of notebook contributors?}

For this research question, we first apply data exploration techniques on the whole dataset for all the features listed in Table \ref{tab:features}. The methods include the frequency count,  mean, variance, outlier tests, and correlation (such as Pearson Correlation) for all pairs. This will give us an overview of the contributed notebooks. Then, we partition the data according to the ranking of the notebook contributors. For each group, we perform the same exploration techniques to identify features that can help distinguish the notebook in each group. We will perform feature-important studies based on the classification task using logistic regression, random forests, or XGBoost. 

\textbf{RQ2:~Do the changing of notebook characteristics correspond to the notebook contributor ranking progression?}

\fix{R:R1.5}\rev{To answer this research question, we leverage the fact that notebooks are labeled based on the chronological rank progression of contributors. This enables us to observe whether the changes in feature values across contributor ranks are significant. We can utilize the features listed in Table \ref{tab:features} for our analysis. Specifically, we first identify contributors who created notebooks while they were at various rank levels. Then, for each individual contributor's ranking change, we can apply statistical techniques to investigate whether the characteristics of notebooks differ when they advance in rank.}
% , we first identify the contributors that change ranks. Then, we tagged their \rev{notebooks} based on their ranking at the time of \rev{notebooks} creation. 
% We then utilize the features listed in Table \ref{tab:features} for our analysis. Since the notebooks are labeled based on the chronological rank progression of contributors, we were able to observe whether the changing feature values across contributor ranks were significant.
% For the ranking change for each individual contributor, we can use the analysis method in RQ1 to identify the features that change from one rank to the next. 
% we utilize the features listed in Table 2 for our analysis. Since the notebooks are labeled based on the chronological rank progression of contributors, we were able to observe whether the changing feature values across contributor ranks were significant.
In addition, we can analyse the trend of each feature throughout the ranking change using regression. We can validate this by creating the classification task based on the key features. After we train a prediction model, we can ask if a contributor changes its values on the identified features following our results, i.e., can we correctly predict if a contributor will be promoted to a higher rank?  

\textbf{RQ3:~Do the characteristics of grandmaster’s notebooks correlate with the highly-popular notebooks in GitHub projects?}

We answer this RQ by applying the study performed in RQ1 to open-source data science projects in GitHub. We will collect the GitHub project as explained in Section~\ref{sec:github}. 
\rev{After retrieving the set of both \rev{highly-popular and unpopular} data science projects based on the statistical distribution of their popularity (i.e., stars and forks)}, we will extract the features (Table~\ref{tab:features}), except the Kaggle's specific features, including Tier, Dataset, and Tag. 
While we use the contributor's tier in Kaggle notebooks, we will use the number of stars and forks to be proxies for \rev{the quality of data science GitHub projects.}
Then, we will perform the statistical test among each feature of notebooks from Kaggle and GitHub. 

\fix{R:R2.10}\rev{For the highly-popular Kaggle grandmaster notebooks and data science GitHub projects, we define the null hypothesis as $H_0$: there is no difference between the feature of the grandmaster's notebooks and highly-popular GitHub data science project notebooks.
For the unpopular Kaggle grandmaster notebooks and data science GitHub projects, we define the null hypothesis as $H_0$: 
there is no difference between the feature of the low-ranking notebooks and unpopular GitHub data science project notebooks. Then, we test the two hypotheses by performing a statistical test of the data from each feature between the Kaggle grandmaster notebooks and GitHub data science projects by following the guidelines by du Prel et al.~\cite{DuPrel2010}. First, in the case of the features that are continuous, we will choose between the t-test and the Mann-Whitney U test depending on the normality of the data. We will use the level of significance ($\alpha$) at
0.05. Second, in the case of the features that are categorical, we will choose the chi-square test. We will use the same $\alpha$ at 0.05.}
We also plan to compare the set of deterministic features to determine the notebook's popularity between Kaggle and GitHub.
%Lastly, we will collect the results and discuss the findings. \fix{R:R1.6}\fix{R:R1.7}

\section{Implications and Impact}
\label{section:impact}
This study has the following implications on both the researchers and practitioners.

\textbf{For researchers.}
The findings from this study aid the understanding of researchers on the important characteristics to determine good computational notebooks. 
For instance, answering RQ1 can provide concrete features that can improve the quality of a notebook.
This can lead to automation, e.g., an automated tool or collection of best practices. Moreover, the findings of important characteristics can also be used to select criteria for Jupyter Notebooks in future studies.
Additionally, RQ3 studies the grandmaster by comparing notebook quality against traditional metrics.
Due to the documentation flexibility of notebooks, the results do provide guidelines on the relationship between documentation and executable code, cross-cutting the field of software documentation and code analysis.

\textbf{For practitioners and developers.} The findings (RQ2) from this study can be used as guidelines for practitioners and developers when developing their Jupyter Notebooks. 
In terms of learning, our findings can lead novice practitioners to improve their skills by focusing on the identified important characteristics. 
%The practitioners can put their focus on the important features identified from the grand master's notebooks to 

\section{conclusion and Future Work}
\label{conclusion}
We present our exploratory study of mining Jupyter Notebooks on Kaggle and GitHub. We plan to apply statistical and data analysis techniques to the four aspects of extracted features to identify characteristics influencing the quality of the notebooks.
For future work, based on our findings, we will create a training guideline for data science practitioners and developers. 
This guideline will help them to improve the quality of the notebooks and advance through the contribution ranking on Kaggle. 
Then, we will conduct an evaluation of the effectiveness of this guideline using a user study.

\section*{Acknowledgment}
This research project is supported by Mahidol University. 

\bibliographystyle{ACM-Reference-Format}
\bibliography{reference}
\end{document}